\documentstyle[preprint,prl,aps,epsf,floats]{revtex}
\tightenlines

\newcommand{\beq}{\begin{equation}}
\newcommand{\eeq}{\end{equation}}
\newcommand{\bqa}{\begin{eqnarray}}
\newcommand{\eqa}{\end{eqnarray}}

\newcommand{\Log}[2]{ \log{ {{#1}+{#2}\over{#1}-{#2}} } }
\newcommand{\LogTwo}[2]{ {\rm log}^2 {{#1}+{#2}\over{#1}-{#2}}  }

\def\square{\vcenter{\vbox{\hrule height.4pt
          \hbox{\vrule width.4pt height8pt
          \kern8pt\vrule width.4pt}\hrule height.4pt}}}

\def\sumint{\hbox{$\sum$}\!\!\!\!\!\!\int}

\begin{document}

\preprint{
\vbox{\halign{&##\hfil\cr
        & hep-ph/9908323  \cr
&\today\cr }}}

\title{ Hard-thermal-loop Resummation \\
                of the Free Energy of a Hot Quark-Gluon Plasma }

\author{Jens O. Andersen, Eric Braaten, and Michael Strickland\footnote{Currently at:  Physics Department, University of Washington, Seattle  WA 98195-1560}}
\address{Physics Department, Ohio State University, Columbus OH 43210, USA}

\maketitle

\begin{abstract}
The quark contribution to the 
free energy of a hot quark-gluon plasma is calculated to leading order
in hard-thermal-loop (HTL) perturbation theory. 
This method selectively resums higher order corrections 
associated with plasma effects,
such as
screening, quasiparticles, and Landau damping.  
Comparing to
the weak-coupling expansion  of QCD, the error in the one-loop HTL free energy
is of order $\alpha_s$, but the large $\alpha_s^{3/2}$
correction from QCD plasma effects is included exactly.

\end{abstract}
\draft
\newpage
\section{Introduction}

Experimental data from ultrarelativistic heavy-ion collisions 
at RHIC will soon 
become available. In order to determine if a quark-gluon plasma has been 
created, a careful comparison of the predictions of hadronic models
and QCD has to be made.
It is therefore desirable to find a systematic way to calculate 
the thermodynamic properties and 
signatures
of a quark-gluon plasma within QCD. Asymptotic freedom suggests that 
at sufficiently high temperatures
a 
straightforward perturbative expansion should suffice. However, at 
experimentally accessible temperatures, perturbative QCD does not seem to be of
any quantitative use [1-3].
The problem is evident
in the free energy of the quark-gluon plasma. The weak-coupling expansion
has been calculated through order $\alpha_s^{5/2}$ [1-3].
The successive approximations to the free energy 
show no sign of converging at temperatures that 
are relevant
for heavy-ion collisions.  

One possibility for improving the convergence of the
perturbative predictions is to apply Pad\'e approximants to the series in
$\alpha_s^{1/2}$~\cite{Pade}, however, this technique can only be applied 
if several terms in the perturbation series are known.  Another possibility
is to use hard-thermal-loop (HTL) approximations within a self-consistent
$\Phi$-derivable framework~\cite{rebhan}.
A third approach is to apply HTL perturbation theory~\cite{EJM1}, 
which is an extension of the resummation method of
Braaten and Pisarski~\cite{Braaten-Pisarski}
into a systematic perturbative expansion.
In two previous papers~\cite{EJM1},
we calculated the one-loop free energy of pure-glue QCD using
HTL perturbation theory.
Here we extend that calculation
to include quarks, thus 
completing the calculation of the free energy of a hot quark-gluon plasma
to leading order in HTL perturbation theory.

The paper is organized as follows. In the next section, we calculate the 
quark contribution to the
free energy of a hot quark-gluon plasma at one loop in HTL
perturbation theory. In section III, we carry out the
 high-temperature expansion
of the free energy, and in section IV we take the low-temperature limit.
In section V, we compare the one-loop HTL free energy with the
weak-coupling expansion of QCD. We conclude in section VI. 
In the
appendix, we have collected the integrals that are required in the calculations.
\section{Quark Contribution to HTL Free Energy}
The one-loop HTL free energy for an $SU(N_c)$ gauge theory with
$N_f$ massless quarks is
\bqa
\label{freedef}
{\cal F}_{\rm HTL} = (N_c^2-1)\left[(d-1){\cal F}_T+{\cal F}_L\right]
	+N_c N_f{\cal F}_q+\Delta{\cal F} \, ,
\eqa
%
where ${\cal F}_T$ and ${\cal F}_L$ are the contributions to the free energy
from transverse and longitudinal gluons, respectively,
${\cal F}_q$ is the contribution to the free energy from 
each flavor and color of the quarks, and 
$\Delta{\cal F}$ is a counterterm.  The quark contribution is given
by 
\bqa
{\cal F}_{q} &=& -\sumint_K \log \det[K\!\!\!\!/-\Sigma(K)]\,.
\label{q-def}
\eqa
The sum-integral in~({\ref{q-def}) represents a dimensionally regularized integral over the
momentum ${\bf k}$ and a sum over the Matsubara frequencies 
$\omega_n=(2n+1)\pi T$:
\bqa
\sumint_K\equiv T \sum_{n=-\infty}^{\infty}\mu^{3-d}\int{d^d k\over(2 \pi)^d}
\,.
\eqa
The factor of $\mu^{3-d}$, where $\mu$ is a renormalization scale,
ensures that 
the regularized free energy has the
correct dimensions even for $d\neq 3$. 
The HTL quark self-energy in~(\ref{q-def}) is
\bqa
\Sigma(K)={m_q^2\over2k}\Log{i\omega_n}{k}\gamma_0
+{m_q^2 \over k}
\left(1-{i\omega_n\over2k} \Log{i\omega_n}{k} \right) 
\hat{\bf k}\cdot\mbox{\boldmath $\gamma$}\, ,
\eqa
where $\hat{\bf k} = {\bf k}/k$ and $m_q$ is the thermal quark mass 
parameter~\cite{Klimov-Weldon}.

We proceed to calculate the quark contribution
to the HTL free energy.
The inverse quark propagator can be written as
\beq
\label{invpr}
K\!\!\!\!/-\Sigma(K)= A_0(K) \gamma_0 - A_S(K) {\hat{\bf k}}
\cdot\mbox{\boldmath $\gamma$}
\, ,
\eeq
%
where
\bqa
\label{a0}
A_0(K) &=& i\omega_n - {m_q^2 \over 2 k} \Log{i\omega_n}{k} \, , \\
\label{as}
A_S(K) &=& k + {m_q^2 \over k} \left( 1 - {i \omega_n \over 2k}  \Log{i\omega_n}{k} \right) \, .
\eqa
We can write the quark contribution~(\ref{q-def}) as
\bqa
\label{sepout}
{\cal F}_{q} &=& -2\sumint_K\log (k^2+\omega_n^2)
-2\;\sumint_K \log \left[{A_S^2-A_0^2\over k^2+\omega_n^2}\right] \, ,
\eqa
where we have separated out the free energy of an ideal gas of
massless fermions.
The first sum-integral in~(\ref{sepout}) can be evaluated analytically.
In the second sum-integral, the sum over Matsubara frequencies 
can be expressed as a contour integral:
\bqa
\label{freequark}
{\cal F}_q = -{7\pi^2\over180}T^4+
2\int_{{\bf k}}\int_C{d\omega\over2\pi i}
	\log\left[{(A_0-A_S)(A_0+A_S)\over \omega^2-k^2}\right]{1\over e^{\beta\omega}+1}\, ,
\eqa
where the contour $C$ 
encloses the points $\omega=i\omega_n$
along the imaginary axis in Fig.~\ref{quarkcon}.
We have introduced a condensed notation for the 
dimensionally regularized momentum integral:
\bqa
\int_{\bf k}=\mu^{3-d}\int{d^d k\over(2 \pi)^d}.
\eqa

\begin{figure}[htb]
\begin{center}
\hspace{1cm}
\epsfysize=8cm
\centerline{\epsffile{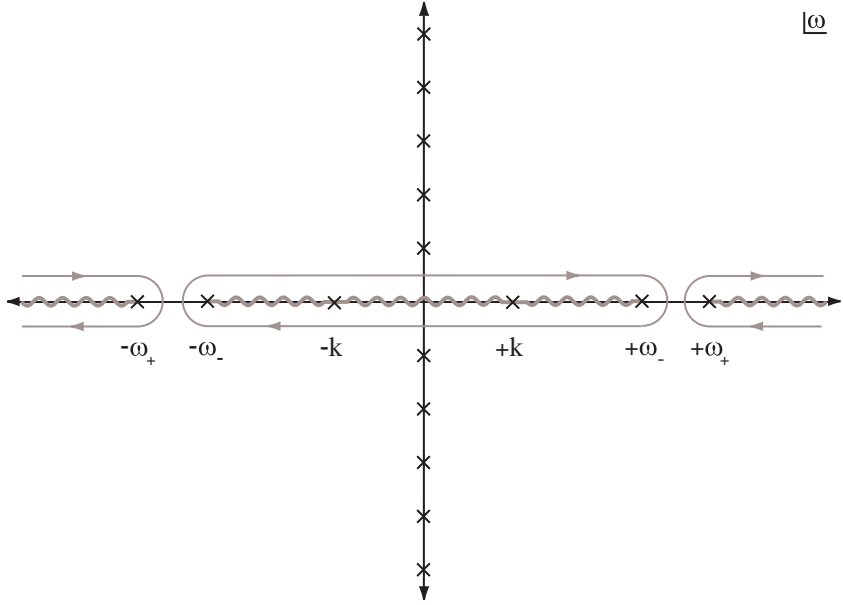}}
\vspace*{0.1cm}
\caption[a]{The quark contribution to the HTL
free energy can be expressed as an
integral over a contour $C$ 
that wraps around
the branch cuts of $\log[(A_0^2-A_S^2)/(\omega^2-k^2)]$.}
\label{quarkcon}
\end{center}
\end{figure}

The integrand in~(\ref{freequark}) has logarithmic 
branch cuts that run from $-\infty$ to $-\omega_+$,
from $-\omega_-$ to $\omega_-$,
from $\omega_+$ to $+\infty$, and from $-k$ to $k$, where $\omega_{\pm}$ 
are the quasiparticle dispersion relations that satisfy
$A_0\mp A_S=0$, or
\bqa
0 &=& \omega_{\pm} \mp k - {m_q^2\over2k}\left[ \left(1\mp{\omega_{\pm}\over k}\right) \Log{\omega_{\pm}}{k} \pm 2\right]\, .
\eqa
%
$\omega_+$ is the dispersion relation for 
the standard quark mode whose helicity equals its chirality.
$\omega_-$ is the dispersion relation for the 
{\it plasmino}, a collective mode
whose helicity is opposite to its chirality~\cite{Klimov-Weldon}.
The integrand in~(\ref{freequark}) 
also has branch cuts running from $-k$ to $k$ due
to the logarithms in~(\ref{a0}) and~(\ref{as}).
The contour $C$ can be deformed to wrap around the branch cuts as shown 
in Fig.~\ref{quarkcon}.
We identify the contributions from the branch cuts that end at $\pm\omega_+$
as the quasiparticle contribution to ${\cal F}_q$ from the quark mode.
We identify the contribution from $k<|\omega|<\omega_-$ as the 
quasiparticle contribution from the plasmino.
The sum of these contributions is denoted by ${\cal F}_{q,\rm qp}$:
\bqa
\label{qp}
{\cal F}_{q,\rm qp} = -4\int_{{\bf k}}\left[T\log(1+e^{-\beta\omega_+})+{1\over2}\omega_{+}
	+T\log{1+e^{-\beta\omega_-}\over 1+e^{-\beta k}} +{1\over2}(\omega_--k)\right] \, .
\eqa
%
We identify the remaining contribution from $|\omega|<k$ as the Landau-damping
term and denote it by ${\cal F}_{q,\rm Ld}$:
\bqa
\label{Ld}
{\cal F}_{q,{\rm Ld}}={4\over\pi}\int_{\bf k}\int_{0}^{k}d\omega\;\theta_q
	\left[{1\over e^{\beta\omega}+1}-{1\over2}\right] \, .
\eqa
%
The angle $\theta_q$ is
\bqa
\theta_q(\omega,k) &=& \arctan{{\pi m_q^4\over k^2}\left[{\omega\over k}+{K^2\over2k^2}L\right]
	\over K^2+2m^2_q +{m_q^4\over k^2}\left[1-{\omega\over k}L-{K^2\over4k^2}(L^2-\pi^2)\right]} \, ,
\eqa
%
where $L=\log[(k+\omega)/(k-\omega)]$ and $K^2=k^2-\omega^2$.
The complete quark contribution to the 
free energy is the sum of the quasiparticle 
term~(\ref{qp}) and the Landau-damping term~(\ref{Ld}). 

We first simplify the quasiparticle term.
The integral of $\omega_+$ is ultraviolet divergent since 
the asymptotic behavior of the dispersion relation
$\omega_+$ is~\cite{pisarski1}
\bqa
\omega_+(k)\longrightarrow
k+{m^2_q\over k}-{m_q^4\over2k^3}\log\left({2k^2\over m^2_q}\right)\, .
\eqa
%
The integral of $(\omega_--k)$ is convergent because 
the dispersion relation $\omega_-(k)$ approaches the light-cone
exponentially fast as $k\to \infty$.
In order to extract the divergence analytically, we make a subtraction that
renders the integral finite in $d=3$ dimensions. 
The subtraction is then evaluated analytically using dimensional
regularization. Our choice of subtraction integral for
the quasiparticle term is
\bqa
\label{qpsub0}
{\cal F}_{q,\rm qp}^{(\rm sub)} = - 2\int_{\bf k}
	\left[\sqrt{k^2+2m_q^2} -{m^4_q\over2(k^2+2m^2_q)^{3/2}}
	\left(\log{2(k^2+2m^2_q)\over m^2_q}-1\right) \right] \, .
\eqa
%
After subtracting this term from~(\ref{qp}), we can take the limit
$d\to3$:
\bqa\nonumber
{\cal F}_{\rm qp}-{\cal F}_{\rm qp}^{(\rm sub)}&=&-
	{2T\over\pi^2}\int_0^{\infty}dk\;k^2\left[\log(1+e^{-\beta\omega_+})
	+\log{1+e^{-\beta\omega_-}\over 1+e^{-\beta k}}\right]\\ \nonumber
&&\hspace{-2.12cm} \null - {1\over\pi^2}\int_0^{\infty}dk\;k^2\left[
	\omega_+- \sqrt{k^2+2m_q^2}+{m^4_q\over2(k^2+2m^2_q)^{3/2}}
	\left(\log{2(k^2+2m^2_q)\over m^2_q}-1\right)\right]\\
\label{qpfin}
&&\null\hspace{-2.12cm}-{1\over\pi^2}\int_0^{\infty} dk\;k^2(\omega_--k)\, .
\eqa
%
If we impose a momentum cutoff $k<\Lambda$, our subtraction 
integral~(\ref{qpsub0}) has power divergences proportional to $\Lambda^4$
and $m_q^2\Lambda^2$ and logarithmic divergences proportional to 
$m_q^4\log\Lambda$ and $m_q^4\log^2\Lambda$.
The quartic divergence is cancelled by the usual renormalization of the
vacuum energy density at zero temperature.
Dimensional regularization throws away the power divergences and replaces
the logarithmic divergences by poles in $d-3$. 
In the limit $d\to3$, the individual integrals in~(\ref{qpsub0}) 
are given by~(\ref{qpsub1})--(\ref{qpsub3}) in the appendix. The result is
\bqa
\label{qpsub}
{\cal F}_{q,\rm qp}^{(\rm sub)}= {1\over2}m_q^4\left({2m_q^2\over\mu^2}\right)^{-\epsilon}
	{\Omega_d\over(2\pi)^d}\left[{1\over\epsilon^2}+{2\log2\over\epsilon}
	-{5\over2}+2\log^22+{\pi^2\over6}\right] \, ,
\eqa
where $d=3-2\epsilon$ and $\Omega_d=2\pi^{d/2}/\Gamma(d/2)$.
The last two integrals in~(\ref{qpfin}) are functions of $m_q$ only
and must therefore be proportional to $m_q^4$. Calculating the integrals
numerically, their contributions to~(\ref{qpfin}) are
$(2.166\times10^{-2})m_q^4$ and $(-1.267\times 10^{-2})m_q^4$, respectively.

We next simplify the Landau-damping term~(\ref{Ld}).
The temperature-independent integral has ultraviolet divergences 
from the region $k\to\infty$ with $\omega\sim k$.
We must again isolate the divergences by making subtractions.
Our choice for the subtraction integral is
\bqa
\label{Ldsub}
{\cal F}_{q,\rm Ld}^{\rm (sub)} = - 2m_q^4 \int_{\bf k}\int_{0}^{k}d\omega
	\left[ {\omega \over k^3(K^2+2m_q^2)} + {K^2 L\over 2k^4(K^2+2m^2_q)}
\right] \, .
\eqa
Subtracting this from~(\ref{Ld}), we can take the limit $d\to3$:
\bqa\nonumber
{\cal F}_{q,\rm Ld}-{\cal F}_{q,\rm Ld}^{\rm (sub)}&=&
	{2\over\pi^3}\int_{0}^{\infty}d\omega{1\over{e^{\beta\omega}+1}}\int_{\omega}^{\infty}dk\;k^2 \theta_q\\
\label{Ldfin}
&& \null - {1\over\pi^3}\int_{0}^{\infty}d\omega\int_{\omega}^{\infty}dk\;k^2
	\left[\theta_q - {\pi m^4_q\omega\over k^3(K^2+2m_q^2)} - 
{\pi m^4_qK^2L\over2k^4(K^2+2m^2_q)}\right] \, .
\eqa
%
If we impose ultraviolet
cutoffs $\omega<\Lambda$ and $k<\Lambda$ on the energy and
momentum, the subtraction integral~(\ref{Ldsub}) has logarithmic
divergences proportional to $m_q^4\log\Lambda$
and $m_q^4\log^2\Lambda$.
They 
cancel against the corresponding divergences
in the quasiparticle subtraction
integral~(\ref{qpsub0}). 
The subtraction integral in~(\ref{Ldsub}) is evaluated in the limit $d\to3$
using~(\ref{ldsub2})--(\ref{ldsub1}):
\bqa \label{Ld-sub}
{\cal F}_{q,\rm Ld}^{\rm (sub)} = - {1\over2}
	m_q^4\left({2m_q^2\over\mu^2}\right)^{-\epsilon}{\Omega_d\over(2\pi)^d}
	\left[{1\over\epsilon^2}+{2\log2\over\epsilon}-4\log2+2\log^22+{\pi^2\over6}\right] \, .
\eqa
The last integral in~(\ref{Ldfin}) is a function of $m_q$ only and is therefore
proportional to $m_q^4$. Its contribution to ~(\ref{Ldfin}) is
$(7.525\times 10^{-3})m_q^4$.

Adding~(\ref{qpfin}), ~(\ref{qpsub}), ~(\ref{Ldfin}) and ~(\ref{Ld-sub}), 
our final result for the quark contribution to the HTL free energy is
\bqa \nonumber 
{\cal F}_q &=& -{2T\over\pi^2}\int_0^{\infty}dk\;k^2\left[\log(1+e^{-\beta\omega_+})
	+\log{1+e^{-\beta\omega_-}\over1+e^{-\beta k}} \right]
	\\
&& 
\label{quarkfin}
\null+ {2\over\pi^3}\int_0^{\infty}d\omega\;{1\over e^{\beta\omega}+1}
	\int_{\omega}^{\infty}dk\;k^2\theta_q \;+\;
\left(2.342\times 10^{-2}\right)m_q^4 \, .
\eqa
Since the quark contribution has no logarithmic ultraviolet divergences, the
counterterm 
$\Delta{\cal F}$ in~(\ref{freedef}) 
is the same as in the pure-glue case~\cite{EJM1}.
If we had used a momentum cutoff $\Lambda$ instead of dimensional 
regularization, we would need
a counterterm 
proportional to $m_q^2\Lambda^2$ to
cancel the quadratic divergence from the quasiparticle term~(\ref{qp}).
\section{High-temperature Expansion}
If the temperature is much larger than the quark mass parameter $m_q$,
the quark contribution to the free energy can be expanded in powers
of $m_q/T$. The integral in~(\ref{freequark})
involves two energy and momentum scales: 
the ``hard'' scale $T$ and the ``soft''
scale $m_q$. The terms in the high-temperature expansion can  receive
contributions from both scales. Dimensional regularization makes it
easy to separate these contributions. 
The soft contribution is obtained by expanding the statistical factor
$1/(e^{\beta\omega}+1)$ in~(\ref{freequark}) in powers of $\omega/T$.
Using the methods in~\cite{EJM1},
one can show that the soft contribution to ${\cal F}_q$ vanishes with
dimensional regularization.
The hard contribution is obtained by expanding the logarithm 
in~(\ref{freequark}) in powers of $m_q^2$.  
The first term in this expansion is
\bqa
{\cal F}_q^{(1)}=  - 4m_q^2
\int_{{\bf k}}\int_C{d\omega\over2\pi i}{1\over\omega^2-k^2}{1\over e^{\beta\omega}+1}\, .
\eqa
%
The integrand has single poles at $\omega=\pm k$ and can be evaluated
using the residue theorem. 
The momentum integrals can be evaluated analytically and 
in the limit $d\to 3$ we obtain
\bqa
\label{f1f}
{\cal F}_q^{(1)}={1\over6}m_q^2T^2\, .
\eqa
%
The second term in the expansion is
\bqa
\label{f2}
{\cal F}_q^{(2)} &=& -2 m_q^4 \int_{{\bf k}}\int_C{d\omega\over2\pi i}
	\left[{2\over(\omega^2-k^2)^2} +{1\over k^2(\omega^2-k^2)} \right. \nonumber \\
&& 
\left. \null - {\omega\over k^3(\omega^2-k^2)}\Log{\omega}{k}
	+{1\over 4 k^4}\LogTwo{\omega}{k} \right] {1\over e^{\beta\omega}+1} \, .
\eqa
Using the
residue theorem and collapsing the contour onto the branch cuts
from the logarithms,~(\ref{f2})
reduces to
\bqa
{\cal F}^{(2)}&=&2m_q^4\int_{\bf k}\Bigg\{{1\over k^2}
{\partial\over\partial k}\left({1\over e^{\beta k}+1}\right)-
{1\over k^4}\int_0^{k}d\omega\;\omega
{\partial\over\partial\omega}\left({1\over e^{\beta\omega}+1}\right)
\log{k+\omega\over k-\omega}\Bigg\}\, .
\eqa
The double integral can be evaluated by first integrating over $k$
and then over $\omega$. Expanding around $\epsilon=0$ and keeping terms
only through $\epsilon^0$, we obtain the finite result
\bqa
\label{f2f}
{\cal F}_q^{(2)} &=& {2\log2-1\over 2 \pi^2}m_q^4\, .
\eqa
The final result for the high-temperature expansion through order
$m_q^4$ is the sum of the first term 
in~(\ref{freequark}) and the terms~(\ref{f1f}) and~(\ref{f2f}):
\bqa
\label{highfin}
{\cal F}_q\rightarrow -{7\pi^2\over180}T^4+ {1\over6}m^2_qT^2
+{2\log2-1\over2\pi^2}m_q^4\, .
\eqa
\section{Low-temperature Limit}
It is useful to understand the behavior of the HTL free energy in the
low-temperature limit where $T\rightarrow 0$ with $m_q$ fixed.
In this limit, ${\cal F}_q$ is proportional to 
$m_q^4$. The coefficient could be extracted directly from the 
final expression~(\ref{quarkfin}) for ${\cal F}_q$, but it  
is simpler 
to compute it from our original expression~(\ref{sepout}) 
for the quark contribution to the free energy.
As $T\to 0$, the sum over 
the discrete Matsubara frequencies $\omega_n=(2n+1)\pi T$
becomes an integral over the 
continuous Euclidean energy $\omega$:
\bqa
\label{low-t}
{\cal F}_q\to-{1\over\pi}\int_{-\infty}^{\infty}
	d\omega\;\int_{\bf k}\,\log\left[{A_S^2-A_0^2\over k^2+\omega^2}\right]
\, .
\eqa
After rescaling the energy $\omega\to k\omega$, we obtain
\bqa
{\cal F}_q = -{2\over\pi}\int_{0}^{\infty}
	d\omega\;\int_{\bf k} \,k\, \log\left[ \left(1 + {m^2_q \over k^2} f(\omega)\right)
	\left(1 + {m^2_q \over k^2} {\bar f}(\omega)\right) \right] \, ,
\eqa
%
where
\bqa
f(\omega) &=& {1 \over 1 + i \omega} +i\left({\pi\over 2}-{\rm arctan}\,\omega \right) 
\eqa
and $\bar{f}$ is the complex conjugate of $f$.
%
Integrating over ${\bf k}$, we obtain
\bqa
{\cal F}_q=-{2\over\pi}m_q^4
\left({m^2_q\over\mu^2}\right)^{d-3}
{\Omega_d\over(2\pi)^d}
	{\Gamma({{d+1\over2}})\Gamma({1-d\over2})\over d+1}
	\int_0^\infty \,d\omega\, \bigg\{\left[f(\omega)\right]^{{d+1\over2}}
	+\left[\bar{f}(\omega)\right]^{{d+1\over2}}\bigg\}\, .
\eqa
Expanding around $d=3$, we get
\bqa\nonumber
{\cal F}_q&=&  {1\over2\pi}m_q^4
\left({m^2_q\over\mu^2}\right)^{-\epsilon}
{\Omega_d\over(2\pi)^d}
\Bigg\{
\left({1\over\epsilon}+{1\over2}\right)\int_0^{\infty}d\omega\;
\left[f^2(\omega)+\bar{f}^{2}(\omega)\right]
\label{lowt}
\\&& \hspace{3.8cm}\null -\int_0^{\infty}d\omega\; 
\left[ f^2(\omega)\log f(\omega)+\bar{f}^{2}(\omega)\log\bar{f}(\omega) \right] \Bigg\}\, .
\eqa
%
The integral of $f^2$ can be 
evaluated analytically 
and is purely imaginary. It is cancelled exactly by the integral of 
$\bar{f}^2$.
This cancellation is 
in accord with the observation that the quark contribution
has no logarithmic ultraviolet divergences.
The last integral in~(\ref{lowt}) 
must be evaluated numerically. The result is
\bqa
{\cal F}_q  
\longrightarrow \left(2.342\times 10^{-2}\right)m_q^4 \, ,
\eqa
which is identical to the $m_q^4$ term in our complete
expression~(\ref{quarkfin}) for
the quark contribution to the free energy.
\section{Comparison with Weak-coupling Expansion}
In this section we present the numerical results for the one-loop HTL 
free energy~(\ref{freedef}) with $N_c=3$ and $N_f=3$.
The quark term is given in~(\ref{quarkfin}). The gluon term is given in 
Ref.~\cite{EJM1}. It depends on the gluon mass parameter $m_g$
and on a renormalization scale $\mu_3$ associated with a logarithmically
divergent integral over the three-momentum.
We use the weak-coupling limits of the gluon and quark mass 
parameters~\cite{Klimov-Weldon}:
\bqa
\label{mg}
m_g^2&=&{2\pi\left(6+N_f\right)\over9}\alpha_s(\mu_4)T^2\,, \\
\label{mq}
m_q^2&=&{2\pi\over3}\alpha_s(\mu_4)T^2\,,
\eqa
where $\mu_4$ is the renormalization scale for the running coupling constant.
We use a parameterization of $\alpha_s(\mu_4)$
that includes the effects of two-loop running:
\bqa
\label{2lrun}
\alpha_s(\mu_4)={4\pi\over\beta_0\bar{L}}\left(
1-{2\beta_1\over\beta_0^2}{\log\bar{L}\over\bar{L}}\right)\, ,
\eqa
where $\beta_0=11-{2\over3}N_f$, $\beta_1=51-{19\over3}N_f$, and
$\bar{L}=\log\left(\mu_4^2/\Lambda^2_{\overline{\mbox{\scriptsize MS}}}\right)$.
For the relation between $\Lambda_{\overline{\mbox{\scriptsize MS}}}$
and the critical temperature $T_c$ for the deconfinement phase transition, 
we use the result $T_c=1.05\,\Lambda_{\overline{\mbox{\scriptsize MS}}}$
calculated for $N_f=4$ flavors of dynamical quarks~\cite{tcnf4}.

The leading-order HTL free energy with $N_f=3$ 
is shown in Fig~\ref{qcdpres}. 
It is scaled by the free energy of an ideal gas of quarks and gluons:
\bqa
{\cal F}_{\rm ideal}=-{8\pi^2\over45}\left(1+{21\over32}N_f\right)T^4.
\eqa
To illustrate the sensitivity to the choices of the renormalization scales,
we take their central values to be $\mu_3=0.717\,m_g$ and  $\mu_4=2\pi T$
and we allow variations by a factor of two. The shaded band indicates the
resulting range in predictions.
The range of ${\cal F}_{\rm HTL}$ comes predominantly from variation in $\mu_4$
at the highest temperatures shown and from variations in $\mu_3$ at the lowest
temperatures shown.  With our expressions from $m_g$ and $\alpha_s$, 
${\cal F}_{\rm HTL}$ diverges either to $+\infty$ or $-\infty$ at $\mu_4 = 
\Lambda_{\overline{\mbox{\scriptsize MS}}}$, depending on whether $\mu_3$ is
greater than or less than our central value of $0.717\,m_g$.
The free energy of an $SU(3)$ gauge theory with $N_f$ massless quarks
has been calculated in the weak-coupling expansion through order 
$\alpha_s^{5/2}$~\cite{Kastening-Zhai,Kastening-Zhai2,Braaten-Nieto}:
\begin{eqnarray}
{\cal F} \;=\; - {8 \pi^2 \over 45} T^4\;
\left[ {\cal F}_0
\;+\; {\cal F}_2  {\alpha_s \over \pi}
\;+\; {\cal F}_3  \left( {\alpha_s\over \pi} \right)^{3/2}
\;+\; {\cal F}_4  \left( {\alpha_s \over \pi} \right)^2
\right.
\nonumber \\
\left.
\;+\; {\cal F}_5  \left( {\alpha_s \over \pi} \right)^{5/2}
\;+\; O(\alpha_s^3 \log \alpha_s) \right] \,.
\label{freeg}
\end{eqnarray}
%
The coefficients in this expansion with $\mu_4=2\pi T$ are
\begin{eqnarray}
{\cal F}_0 &=& 1 + \textstyle{21\over 32}N_f  \, ,
\\
\label{correct}
{\cal F}_2 &=& - {15 \over 4} \left( 1 + \textstyle{5\over 12}N_f  \right)\;,
\\
{\cal F}_3 &=& 30 \left( 1 + \textstyle{1\over 6}N_f \right)^{3/2} \, ,
\\
{\cal F}_4 &=&  237.2 + 15.97 N_f - 0.413 N_f^2
+ { 135 \over 2} \left( 1 + \textstyle{1\over 6}N_f  \right)
	\log \left[ {\alpha_s \over \pi}
		\left(1 + \textstyle{1\over 6}N_f \right) \right]\, , 
\\
{\cal F}_5 &=& -\left( 1 + \textstyle{1 \over 6}N_f\right)^{1/2}
\Bigg[ 799.2 + 21.96 N_f + 1.926 N_f^2\Bigg]\, .
\end{eqnarray}
The predictions from the weak-coupling
expansion with $\mu_4=2\pi T$ are
compared to the HTL free energy in Fig.~\ref{qcdpres}.
The expansions of the pressure
truncated after orders $\alpha_s$, $\alpha_s^{3/2}$, $\alpha_s^2$,
and $\alpha_s^{5/2}$ are shown as the dashed lines
labelled 2, 3, 4, and 5.
As successive terms in the weak-coupling expansion
are added, the predictions fluctuate wildly.
In addition, the sensitivity to the renormalization scale $\mu_4$
increases at each successive order.
Of course, because of asymptotic freedom,
the first few terms in the weak-coupling expansion will appear to 
converge at sufficiently high temperature. However, this occurs only at
enormously high temperatures, where all the corrections to the 
ideal gas are tiny.
For example, for $N_f=3$,
the $\alpha_s^{3/2}$ correction is smaller 
than the $\alpha_s$ correction
only if $\alpha_s<3\pi/128$. 
If we use~(\ref{2lrun}) to extrapolate to high temperature while ignoring
the effects of heavier quark flavors, this corresponds to 
a temperature $T> 500\,T_c$.

We now compare the high-temperature expansion of the HTL free energy 
in~(\ref{highfin}) with
the weak-coupling expansion.
Using the values~(\ref{mg}) and~(\ref{mq}) for the thermal mass parameters,
we find that the $\alpha_s$ correction 
is overincluded by a factor of $12(3+N_f)/(12+5N_f)$.
The $\alpha_s^{3/2}$ correction
is included exactly in the leading HTL result. 
The overincluded $\alpha_s$ correction and the large positive
$\alpha_s^{3/2}$ correction 
combine with higher order corrections in the HTL free energy
to give a negative correction
that rises slowly with $T$ as shown in Fig.~\ref{qcdpres}.
\begin{figure}[htb]
\begin{center}

\epsfysize=8cm
\centerline{\epsffile{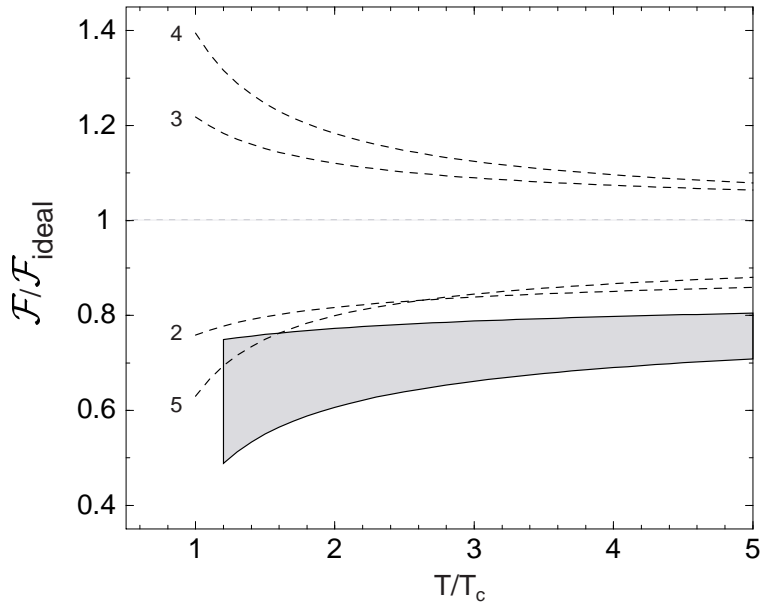}}
\caption[a]{The free energy for QCD with $N_f=3$ quarks as a function of
$T/T_c$. The HTL free energy is shown as a shaded band that corresponds to
varying $\mu_3$ and $\mu_4$ by a factor of two around their central values.
The weak-coupling expansion through orders $\alpha_s$, $\alpha_s^{3/2}$,
$\alpha_s^2$, and $\alpha_s^{5/2}$ are shown as dashed lines
labelled by 
2, 3, 4 and 5.}
\label{qcdpres}
\end{center}
\end{figure}

Lattice gauge theory has been used to calculate the equation of state
of a quark-gluon plasma with $N_f=2$~\cite{nf2,nf2p} and $N_f=4$~\cite{nf4}
flavors of dynamical quarks. These calculations indicate that the pressure,
which is the negative of the free energy, approaches that of an ideal gas
from below. The approach to the ideal gas is
more rapid than the leading order HTL result.
For $N_f=4$, it reaches 80\% of the value for an ideal gas already at 
$T=2.5\,T_c$.
For higher values of $T$, the leading order HTL result lies significantly
below the lattice results.
This is not of great concern, because the difference can be accounted for
by the next-to-leading order correction in HTL perturbation theory.
At next-to-leading order, there are two-loop diagrams and 
one-loop diagrams with HTL counterterms.  The contributions of order 
$\alpha_s$ coming from the hard momentum regions of the two-loop diagrams
will reproduce the order-$\alpha_s$ term in the conventional 
perturbative series (\ref{freeg}).  
The contribution from the HTL counterterm diagram will precisely cancel
the order-$\alpha_s$ term in the one-loop HTL free energy.
Thus the next-to-leading order correction 
to ${\cal F}_{\rm HTL}/{\cal F}_{\rm ideal}$ 
must approach $[10(24+7N_f)/(32+21N_f)]\alpha_s/\pi$
in the limit $\alpha_s\to0$.
This has the correct sign and roughly the
right magnitude to account for the discrepancy with the lattice results.

\section{Conclusions}
We have completed the calculation of  
the free energy of a quark-gluon plasma to leading order
in HTL perturbation theory by calculating the quark
contribution. 
The quark term has a quadratic ultraviolet divergence that vanishes with
dimensional regularization, but it has no
logarithmic
ultraviolet divergences.
Comparing our result to the weak-coupling expansions for the
free energy, we find that the error is of order $\alpha_s$ but the 
large correction proportional to $\alpha_s^{3/2}$ is included exactly.
It is therefore possible that the HTL perturbative expansion
for the free energy will have much better convergence properties than
the conventional weak-coupling expansion. To verify this, it will be
necessary
to extend the calculations of the free energy to next-to-leading order
in HTL perturbation
theory.
\section*{Acknowledgments}
This work was supported in part by the U.~S. Department of 
Energy Division of High Energy Physics (grant DE-FG02-91-ER40690),
by a Faculty Development Grant 
from the Physics Department of the Ohio State University,
by the Norwegian Research Council 
(project 124282/410), 
and by the National Science Foundation (grant PHY-9800964).

\appendix\bigskip\renewcommand{\theequation}{\thesection.\arabic{equation}}
\setcounter{equation}{0}\section{Integrals}
In this appendix, we collect the results for the integrals
that are required to calculate the contribution from quarks to the
one-loop HTL free energy. 
We use dimensional regularization, so that
power ultraviolet divergences are set to zero and logarithmic
ultraviolet divergences appear as poles in $\epsilon$.
In the HTL free energy, the ultraviolet divergences are isolated in
subtraction terms
that must be expanded around $\epsilon=0$ through
order $\epsilon^0$.
The integrals required to evaluate the subtractions
in the quasiparticle terms are
\bqa
\label{qpsub1}
\int_0^{\infty}dk\;k^{2-2\epsilon}\sqrt{k^2+m^2}&=&
-{1\over16}m^{4-2\epsilon}\left[{1\over\epsilon}+2\log2-{1\over2}\right]\, ,\\
\label{qpsub2}
\int_0^{\infty}dk\;k^{2-2\epsilon}{1\over(k^2+m^2)^{3/2}}&=
&{1\over2}m^{-2\epsilon}\left[{1\over\epsilon}+2\log2-2\right]\, ,\\
\label{qpsub3}
\int_0^{\infty}dk\;k^{2-2\epsilon}{1\over(k^2+m^2)^{3/2}}
\log{k^2+m^2\over m^2}&=
&{1\over2}m^{-2\epsilon}\left[{1\over\epsilon^2}-4+{\pi^2\over6}-2\log^22
+4\log2\right]\, .
\eqa
The integrals required to evaluate the subtractions
in the Landau-damping terms are
\bqa
\label{ldsub2}
\int_0^{\infty}d\omega\;\omega\int_{\omega}^{\infty}dk\;k^{-1-2\epsilon}
{1\over k^2-\omega^2+m^2}&=&
{1\over4}m^{-2\epsilon}\left[
{1\over\epsilon^2}+{\pi^2\over6}\right]\, ,\\
\label{ldsub1}
\int_0^{\infty}d\omega\;\int_{\omega}^{\infty}dk\;
k^{-2-2\epsilon}{k^2-\omega^2\over k^2-\omega^2+m^2}\log{k+\omega\over k-\omega}
&=&m^{-2\epsilon}\left[{\log2\over\epsilon}-2\log2+\log^22\right]\, .
\eqa

\end{document}